\documentclass[12pt]{article}
\usepackage{html}
\usepackage{epsf}
\usepackage{graphicx}
\usepackage{amssymb}
\begin{document}
\title{MATTERS OF GRAVITY, The newsletter of the APS Topical Group on 
Gravitation}
\begin{center}
{ \Large {\bf MATTERS OF GRAVITY}}\\ 
\bigskip
\hrule
\medskip
{The newsletter of the Topical Group on Gravitation of the American Physical 
Society}\\
\medskip
{\bf Number 37 \hfill Winter 2011}
\end{center}
\begin{flushleft}
\tableofcontents
\vfill\eject
\section*{\noindent  Editor\hfill}
David Garfinkle\\
\smallskip
Department of Physics
Oakland University
Rochester, MI 48309\\
Phone: (248) 370-3411\\
Internet: 
\htmladdnormallink{\protect {\tt{garfinkl-at-oakland.edu}}}
{mailto:garfinkl@oakland.edu}\\
WWW: \htmladdnormallink
{\protect {\tt{http://www.oakland.edu/?id=10223\&sid=249\#garfinkle}}}
{http://www.oakland.edu/?id=10223&sid=249\#garfinkle}\\

\section*{\noindent  Associate Editor\hfill}
Greg Comer\\
\smallskip
Department of Physics and Center for Fluids at All Scales,\\
St. Louis University,
St. Louis, MO 63103\\
Phone: (314) 977-8432\\
Internet:
\htmladdnormallink{\protect {\tt{comergl-at-slu.edu}}}
{mailto:comergl@slu.edu}\\
WWW: \htmladdnormallink{\protect {\tt{http://www.slu.edu/colleges/AS/physics/profs/comer.html}}}
{http://www.slu.edu//colleges/AS/physics/profs/comer.html}\\
\bigskip
\hfill ISSN: 1527-3431

\bigskip

DISCLAIMER: The opinions expressed in the articles of this newsletter represent
the views of the authors and are not necessarily the views of APS.
The articles in this newsletter are not peer reviewed.

\begin{rawhtml}
<P>
<BR><HR><P>
\end{rawhtml}
\end{flushleft}
\pagebreak
\section*{Editorial}

The next newsletter is due September 1st.  This and all subsequent
issues will be available on the web at
\htmladdnormallink 
{\protect {\tt {https://files.oakland.edu/users/garfinkl/web/mog/}}}
{https://files.oakland.edu/users/garfinkl/web/mog/} 
All issues before number {\bf 28} are available at
\htmladdnormallink {\protect {\tt {http://www.phys.lsu.edu/mog}}}
{http://www.phys.lsu.edu/mog}

Any ideas for topics
that should be covered by the newsletter, should be emailed to me, or 
Greg Comer, or
the relevant correspondent.  Any comments/questions/complaints
about the newsletter should be emailed to me.

A hardcopy of the newsletter is distributed free of charge to the
members of the APS Topical Group on Gravitation upon request (the
default distribution form is via the web) to the secretary of the
Topical Group.  It is considered a lack of etiquette to ask me to mail
you hard copies of the newsletter unless you have exhausted all your
resources to get your copy otherwise.

\hfill David Garfinkle 

\bigbreak

\vspace{-0.8cm}
\parskip=0pt
\section*{Correspondents of Matters of Gravity}
\begin{itemize}
\setlength{\itemsep}{-5pt}
\setlength{\parsep}{0pt}
\item John Friedman and Kip Thorne: Relativistic Astrophysics,
\item Bei-Lok Hu: Quantum Cosmology and Related Topics
\item Veronika Hubeny: String Theory
\item Beverly Berger: News from NSF
\item Luis Lehner: Numerical Relativity
\item Jim Isenberg: Mathematical Relativity
\item Katherine Freese: Cosmology
\item Lee Smolin: Quantum Gravity
\item Cliff Will: Confrontation of Theory with Experiment
\item Peter Bender: Space Experiments
\item Jens Gundlach: Laboratory Experiments
\item Warren Johnson: Resonant Mass Gravitational Wave Detectors
\item David Shoemaker: LIGO Project
\item Stan Whitcomb: Gravitational Wave detection
\item Peter Saulson and Jorge Pullin: former editors, correspondents at large.
\end{itemize}
\section*{Topical Group in Gravitation (GGR) Authorities}
Chair: Steve Detweiler; Chair-Elect: 
Patrick Brady; Vice-Chair: Manuella Campanelli. 
Secretary-Treasurer: Gabriela Gonzalez; Past Chair:  Stan Whitcomb;
Members-at-large:
Frans Pretorius, Larry Ford,
Scott Hughes, Bernard Whiting,
Laura Cadonati, Luis Lehner.
\parskip=10pt

\vfill
\eject

\vfill\eject

\section*{\centerline
{GGR program at the APS meeting in Anaheim, CA}}
\addtocontents{toc}{\protect\medskip}
\addtocontents{toc}{\bf GGR News:}
\addcontentsline{toc}{subsubsection}{
\it GGR program at the APS meeting in Anaheim, CA, by David Garfinkle}
\parskip=3pt
\begin{center}
David Garfinkle, Oakland University
\htmladdnormallink{garfinkl-at-oakland.edu}
{mailto:garfinkl@oakland.edu}
\end{center}

We have an exciting GGR related program at the upcoming APS April meeting in Anaheim, CA.  Our Chair-Elect, Patrick Brady did an 
excellent job of putting together this program.  At the APS meeting there will be several invited sessions of talks sponsored
by the Topical Group in Gravitation (GGR).  

A plenary talk on gravitational physics is:\\

Nergis Mavalvala, QM and Gravity Wave Detection\\

The invited sessions sponsored by GGR are as follows:\\

Black Holes: Nature's Ultimate Spinmeisters\\
(joint with DAP)\\
Jeffrey McClintock, Measuring the Spins of Stellar-Mass Black Holes\\
Francois Foucart, Black hole-Neutron Star Mergers: Effects of the Orientation of the Black Hole Spin\\
Massimo Dotti, Spin-Induced Gravitational Recoil: Effect on the Evolution of Massive Black Holes\\

Neutron Stars: What's the Matter?\\
Zachariah Etienne, Numerical Simulations of Binary Systems with Matter Companions\\
Jocelyn Read, Measuring the Neutron Star Equation of State Using Gravitational Waves from Binary Observations\\
Chad Hannah, Searching for Gravitational Waves from Compact Binary Coalescence\\

Observational Implications of Gravitational Wave Observations\\
Joshua Smith, Exploring the Transient Universe with Gravitational Waves\\
Eric Thrane, Searches for a Stochastic Background of Gravitational Waves\\
Teviet Chreiton, Exploring the Galactic Neutron Star Population with Gravitational Waves\\

Progress in Quantum Gravity\\
Parampreet Singh, Progress in Loop Quantum Gravity\\
Wei Song, Kerr/CFT Correspondence\\
Donald Marolf, Unitarity and Holography in Gravitational Physics\\

Einstein Prize and New Methods for Old Problems in Gravitational Physics\\
Ted Newman, Einstein Prize Talk: Light Cones in Relativity: Real, Complex, and Virtual - with Applications\\
Holger Mueller, Gravitational Redshift, Equivalence Principle, and Matter Waves\\
Fethi Ramazanoglu, Evaporation of Two Dimensional Black Holes\\

Frontiers of Computational Astrophysics and Gravitation\\
(joint with DCOMP)\\
Tanja Bode, Correlated Electromagnetic and Gravitational Waves from Supermassive Black Hole Binary Mergers\\
Bruce Allen, The Einstein@Home Search for New Neutron Stars\\
Matthew Turk, Cosmological Simulations: Capturing the Formation of the First Stars and Galaxies\\

Pipkin Award and Quantum Technologies for Gravitational Wave Detection\\
(joint with GPMFC)\\
Daniel Sigg, Squeezed Light Techniques for Gravitational Wave Detection\\
Thomas Corbitt, Quantum Noise and Opto-mechanics in Advanced Gravitational Wave Detectors\\
Michael Romalis, Francis M. Pipkin Award Talk: Lorentz and CPT Symmetry Tests with Atomic Co-Magnetometers\\

The GGR contributed sessions are as follows:\\

Quantum Aspects of Gravitation\\

Self-force Calculations and Their Implications for Gravitational Physics\\

Spacetime Structure of Binary Black Hole Simulations\\

Characterization and Instrumentation for Gravitational Wave Detection\\
(joint with DAP)\\

Gravitational Astrophysics\\

Post-Newtonian Approximations and Alternative Theories of Gravity\\

Strong Field Gravity: Black Holes, Event Horizons, and Cosmology\\

Numerical Relativity: Black Holes, Neutron Stars, and Accretion Disks\\

Numerical Relativity: Algorithms and Code Development\\

Signal Analysis Methods for Gravitational Wave Detection\\

Gravitational Wave Astronomy\\
(joint with DAP)\\

Numerical Relativity: Binaries with Matter\\

\vfill\eject

\section*{\centerline
{we hear that \dots}}
\addtocontents{toc}{\protect\medskip}
\addcontentsline{toc}{subsubsection}{
\it we hear that \dots , by David Garfinkle}
\parskip=3pt
\begin{center}
David Garfinkle, Oakland University
\htmladdnormallink{garfinkl-at-oakland.edu}
{mailto:garfinkl@oakland.edu}
\end{center}

Ted Newman has won the Einstein Prize.

Patrick Brady, David Brown, Tevian Dray, Luis Lehner, and Nergis Mavalvala have been elected as APS Fellows.

Woei-Chet Lim, Mark Hannam, Thomas Corbitt, and Parampreet Singh have been awarded the the GRG Society's Chandrasekhar prize for best postdoc presentations at
GR19.  Amitai Bin-Nun, Bethan Cropp, Samuel Gralla, Charalampos Markakis, Vivien Raymond, Ian
Morrison, David Sloan, Jan Steinhoff, and Francesca Vidotto have been awarded the GRG Society's Hartle prize for best student presentations at GR19.

Hearty Congratulations!

\section*{\centerline
{100 years ago}}
\addtocontents{toc}{\protect\medskip}
\addcontentsline{toc}{subsubsection}{
\it 100 years ago, by David Garfinkle}
\parskip=3pt
\begin{center}
David Garfinkle, Oakland University
\htmladdnormallink{garfinkl-at-oakland.edu}
{mailto:garfinkl@oakland.edu}
\end{center}

In 1911 Einstein calculated the bending of light in a gravitational field using the principle of equivalence and got half
of the correct answer.  He published this result in a paper called ``Uber den Einflus der Schwercraft auf die Ausbreitung des Lichtes''
(On the Influence of Gravitation on the Propagation of Light).  A translation of this paper in English is available at
\htmladdnormallink
{\protect {\tt{http://www.relativitybook.com/resources/Einstein\_gravity.html}}}
{http://www.relativitybook.com/resources/Einstein\_gravity.html}\\

\vfill\eject

\section*{\centerline
{Endpoint of the Gregory-Laflamme Instability}}
\addtocontents{toc}{\protect\medskip}
\addtocontents{toc}{\bf Research briefs:}
\addcontentsline{toc}{subsubsection}{
\it Endpoint of the Gregory-Laflamme Instability, by Gary Horowitz}
\parskip=3pt
\begin{center}
Gary Horowitz, University of California at Santa Barbara
\htmladdnormallink{gary-at-physics.ucsb.edu}
{mailto:gary@physics.ucsb.edu}
\end{center}

In four spacetime dimensions, black holes are believed to be stable. In $D>4$, this is not always the case. As shown by  Gregory and Laflamme \cite{Gregory:1993vy}, the simplest black hole in five dimensional Kaluza-Klein theory can be unstable. This solution is just the product of a $D=4$ Schwarzschild black hole and a circle and is often called a black string. When the length of the circle is greater than the Schwarzschild radius of the black hole, the solution is unstable to long wavelength perturbations. This has become known as the Gregory-Laflamme (GL) instability.

The outcome of this instability has been a matter of intense debate for almost two decades. In their original paper, Gregory and Laflamme showed that when the black string is unstable, its entropy is less than the entropy of a five dimensional spherical black hole with the same total mass.  They concluded that the horizon would pinch off and form a spherical black hole. At the moment the horizon pinches off, it becomes singular. If this  indeed happens, cosmic censorship would be generically violated in five dimensions. 

In 2003, Choptuik et al \cite{Choptuik:2003qd} attempted a numerical calculation of the evolution of the unstable black string using the full nonlinear Einstein equation. They showed that the horizon forms large spherical black holes connected by thin black strings, but then the code crashed (see Fig. 1). This was puzzling since the connecting strings should themselves be GL unstable. In another development, Maeda and I showed that the horizon could not pinch off in finite affine parameter \cite{Horowitz:2001cz}. This  led us to believe  that the outcome would be a nonuniform black string. While technically correct, our paper turned out to be a red herring. Static nonuniform black strings were subsequently found \cite{Wiseman:2002zc}, but shown not to be the endpoint of the GL instability, since their entropy was less than the uniform solution\footnote{This is the case for $D \le 13$. In higher dimension, the nonuniform string can be the endpoint of the instability \cite{Sorkin:2004qq}.}.  Wald then pointed out that it was possible for the horizon to pinch off in infinite affine parameter but still at a finite (advanced) time as seen from infinity.

\begin{figure}[h!]
\begin{center}
\includegraphics[scale=.5]{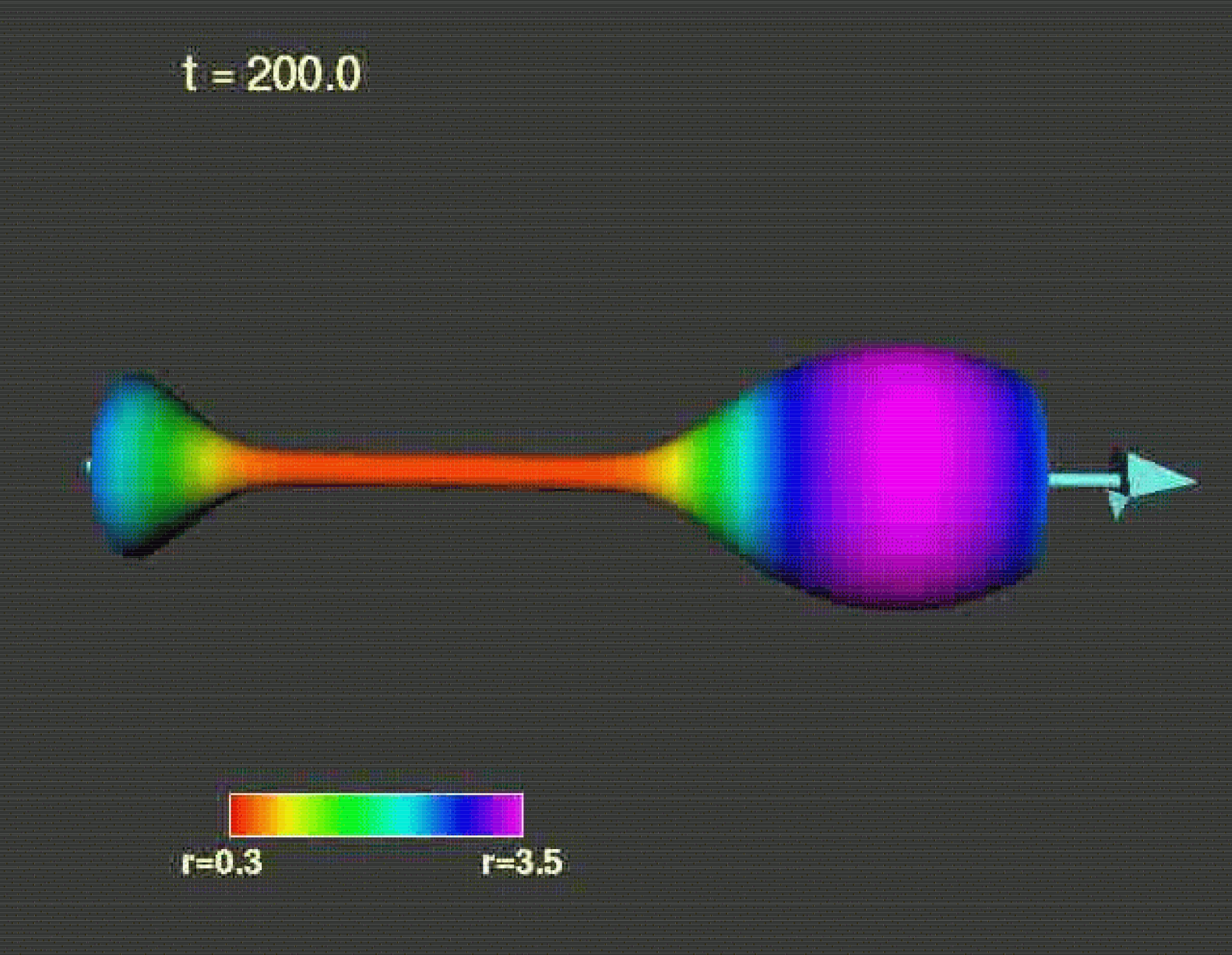}
\end{center}
\vspace{-.5cm}
\caption{
Embedding diagram of the apparent horizon in the original attempt to evolve the unstable black string \cite{Choptuik:2003qd}. Colors denote radius in units of M.
 }
\end{figure}

In a remarkable tour de force, Lehner and Pretorius have recently extended the evolution of the black string much farther than they had before \cite{Lehner:2010pn} \ and discovered a surprising result. The thin black strings connecting the spherical black holes in the earlier evolution are indeed unstable and develop smaller spherical black holes connected by thinner black strings, which are also unstable and form even smaller spherical black holes connected by thinner black strings, etc. (see Fig. 2). They follow this cycle through four generations, but it appears likely to continue indefinitely. There is a form of discrete self similarity in which each generation happens on a smaller scale and in a faster time. Extrapolating to infinite generations, the horizon indeed pinches off in finite time as seen from infinity. In fact, if one stays away from the points that develop spherical black holes, the thickness of the black string goes to zero approximately linearly in time. This is very similiar to the behavior of a fluid when droplets break off.

\begin{figure}[h!]
\begin{center}
\includegraphics[scale=.5]{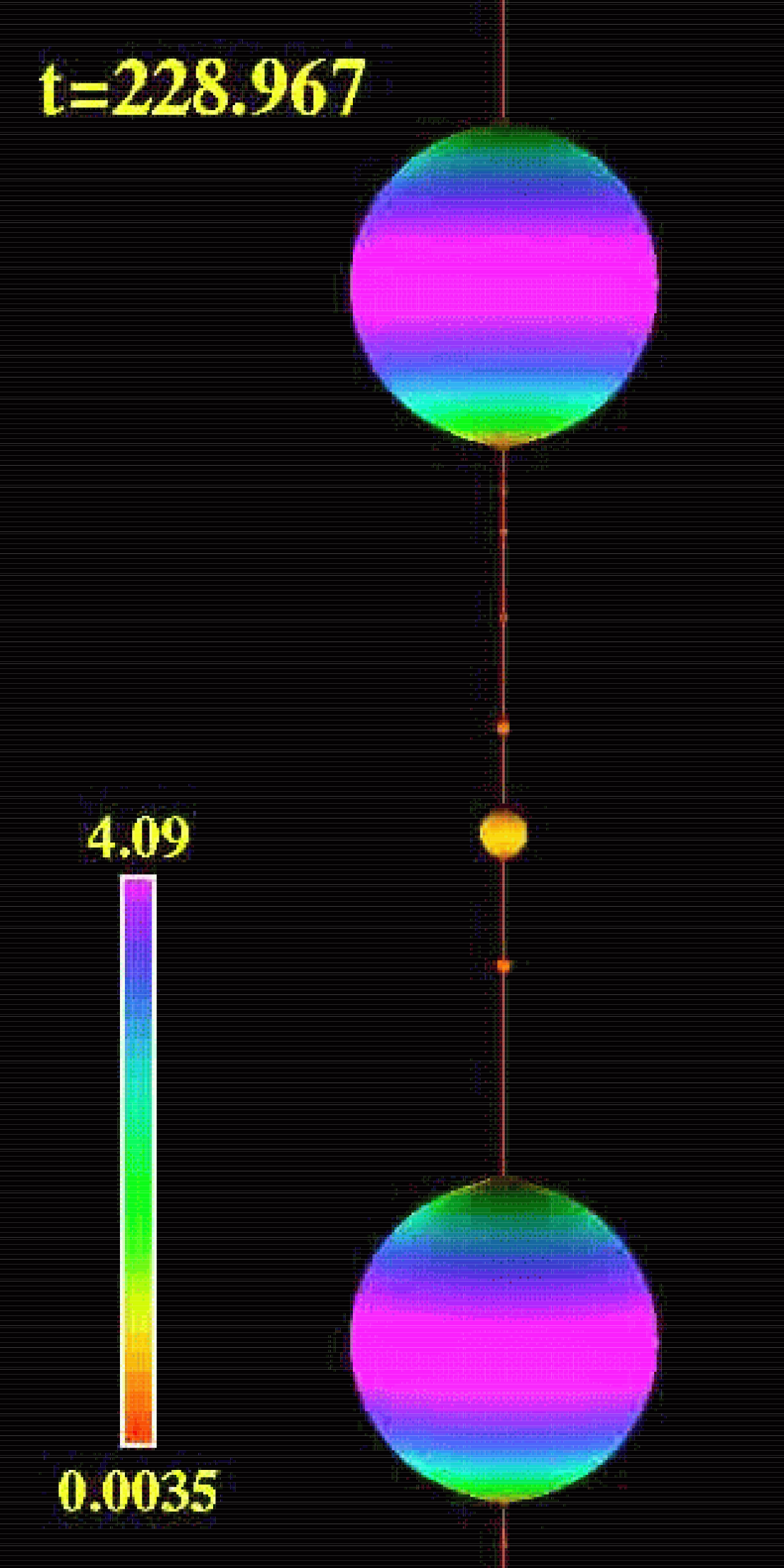}
\includegraphics[scale=.5]{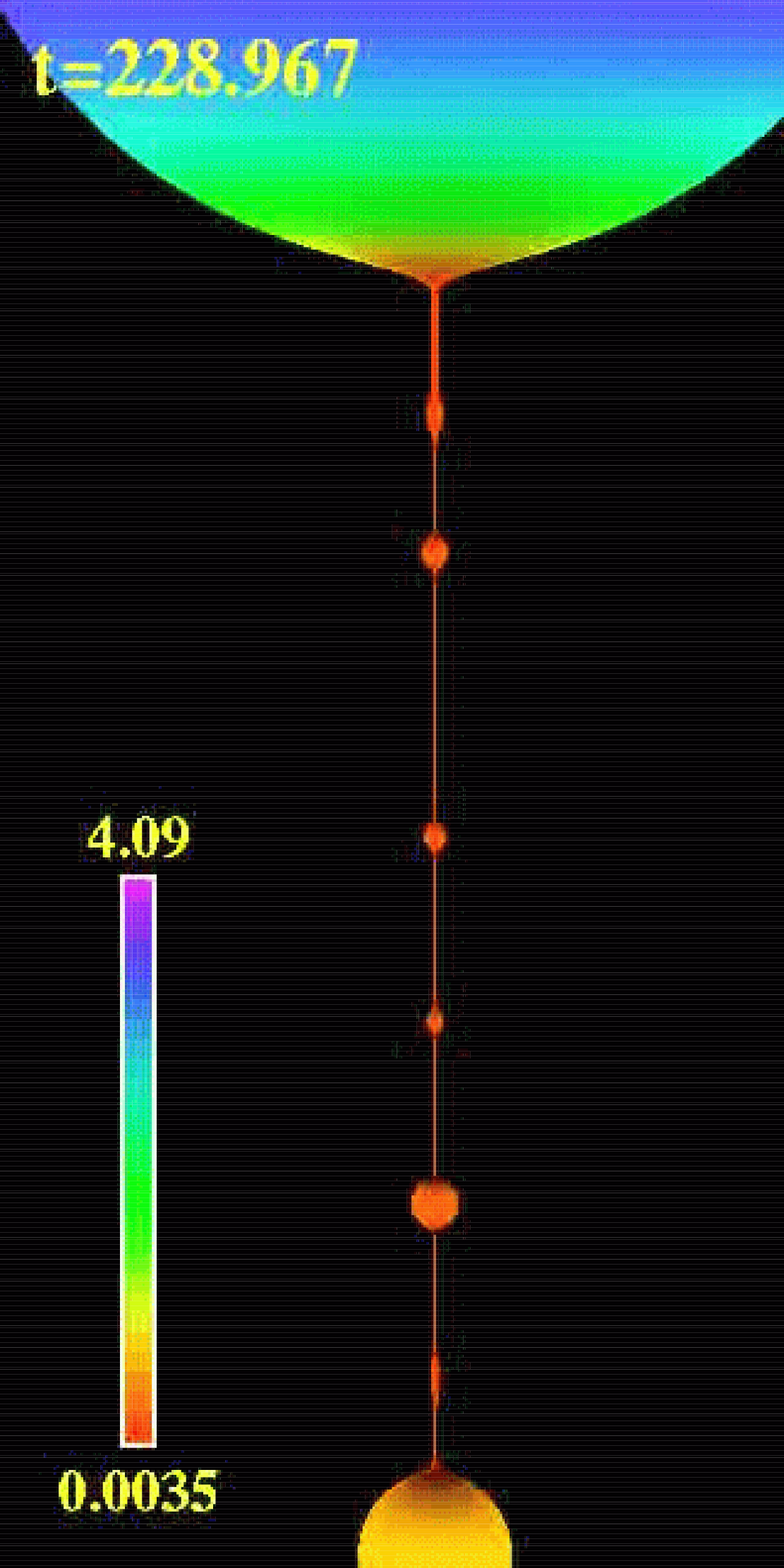}
\end{center}
\vspace{-.5cm}
\caption{
Embedding diagram of the apparent horizon in the new evolution of the unstable black string \cite{Lehner:2010pn}. The figure on the right is a zoomed in version of the one on the left. }
\end{figure}

Thus, it appears that cosmic censorship indeed fails in higher spacetime dimension.  The implications of this remain to be explored. 


\vfill\eject

\section*{\centerline
{Chandrasekhar Symposium}}
\addtocontents{toc}{\protect\medskip}
\addtocontents{toc}{\bf Conference reports:}
\addcontentsline{toc}{subsubsection}{
\it Chandrasekhar Symposium, 
by David Garfinkle}
\parskip=3pt
\begin{center}
David Garfinkle, Oakland University  
\htmladdnormallink{garfinkl-at-oakland.edu}
{mailto:garfinkl@oakland.edu}
\end{center}

The Chandrasekhar Centenial Symposium was held at The University of Chicago, October 16-17, 2010.  Much of 
the content of the symposium is posted on the conference website
\htmladdnormallink
{\protect {\tt{http://physics.uchicago.edu/events/chandra100/}}}
{http://physics.uchicago.edu/events/chandra100/}\\

The conference began with some opening remarks by Bob Wald.  He noted that Chandra had left strict instructions that there were to 
be no ``Memorial Conferences'' for him, and that instead this Chandrasekhar Symposium was to be a celebration of the many areas
of science to which Chandra had contributed.    

Freeman Dyson spoke on ``Chandra's Role in 20th Century Science.''  He noted that Chandra had made a distinction between
``basic science'' (coming up with new theories) and ``derived science'' (working out the consequences of existing theories), and 
that Chandra emphasized that his own work was in derived science.  Dyson pointed to Chandra's work on the white dwarf mass limit as 
a masterpiece of derived science.  He noted how that work became the gateway to further work on the subject of gravitational collapse 
and black hole formation, and especially mentioned Zwicky's work on supernovae and Oppenheimer's work on gravitational collapse.  
It has often been noted that it took a long time for the significance and implications of Chandra's work on the white dwarf mass limit
to be recognized.  Usually this delay is attributed to Eddington; but Dyson pointed out that Zwicky's reputation as an eccentric 
tended to delay the acceptance of his results, and that Oppenheimer's own emphasis on basic science caused him not to follow up or
promote his derived science result on gravitational collapse.  

Cliff Will spoke on ``The Unreasonable Effectiveness of the Post-Newtonian Approximation.'' He began by reminding us that the 
post-Newtonian approximation is essentially an expansion of the Einstein field equation and the matter equations of motion in powers
of $v/c$.  He also mentioned the paramaterized post-Newtonian formalism which allows one to make similar expansions for alternative
theories of gravity.  This makes for a convenient way to find the observational differences between alternative theories and 
general relativity (and by comparison with observation often a convenient way to rule out the alternative theories).  Since 
the post-Newtonian approximation is designed for the regime of low speed and weak gravity, it would be expected to break down 
at high speeds or strong gravity.  But Will noted that it actually works surprisingly well (hence ``unreasonable effectiveness'') 
even in some situations of high speeds and/or strong gravity.  In particular, Will discussed two cases: the binary pulsar, and the
inspiral and merger of binary black holes.  The change in period of the Hulse/Taylor binary pulsar is an indirect detection of 
gravitational radiation through the comparison of observation with the prediction of the post-Newtonian quadrupole formula.  But 
though the neutron stars themselves are slowly moving, their own self-gravity is not weak.  So it is a very welcome and somewhat 
surprising result that the post-Newtonian approximation works well in this situation.  The inspiral and merger of binary black holes 
is a possible candidate for a source of directly detectable gravitational waves, so it is important to model this system to predict
the gravitational waveform that it emits.  The last stages of this process need to be modelled using numerical relativity, and 
the early stages using the post-Newtonian approximation; but what about the intermediate stages where the black holes are no longer
slowly moving but are still far from merger?  Will pointed out that quite surprisingly the post-Newtonian approximation works very
well for these intermediate stages too.   

Roger Penrose spoke on ``Mathematical Properties of Black Holes and Colliding Plane Waves.''  He pointed out that Chandra had 
a major research program (culminating in a book) on the mathematical aspects of black holes and colliding gravitational waves.
Penrose noted that both he and Chandra wanted to understand gravitational collapse, black holes and singularities, but that they
had very different mathematical styles and therefore approached the problem very differently.  Penrose's style is very geometrical, 
so his approach involved proving theorems, using the methods of differential geometry, on the existence of singularities and the
properties of black holes.  In contrast, Chandra's style is very algebraic, so his approach involved studying the properties of
exact solutions and their perturbations.  These exact solutions included the Schwarzschild, Reissner-Nordstrom, and Kerr solutions
for black holes; but they also included colliding plane waves, with an emphasis on the singularities formed in the collision.  
Penrose ended with more speculative ideas on explaining the big bang singularity, and in particular its difference from the 
singularities formed in gravitational collapse.     

Jayant Narlikar spoke on ``Chandra's Impact on Indian Astronomy.'' He mentioned a meeting between Chandra and Saha in 1930, and
talked about an effort to recruit Chandra to return to India.  Though Chandra remained in Chicago, Narlikar noted that there were
many interactions between Chandra and other Indian scientists, and that over all Chandra had a great influence on Indian astronomy. 

John Friedman spoke on ``Instabilities of Relativistic Stars.'' He noted that he and Chandra had worked on the instability of
general relativistic rotating stars, and pointed out that this subject remains one of high interest, especially since some of 
the instabilities involve the emission of gravitational waves.  Friedman mentioned that instabilities place an upper limit on
the rate of rotation for a neutron star.  He listed several different perturbative modes that can lead to instability, and emphasized 
the r-modes as a fruitful area for theoretical investigations.   

Kip Thorne spoke on ``Black Holes.''  He pointed out that numerical relativity simulations give detailed information about the 
inspiral and merger of binary black hole systems.  However, one would like an intuitive understanding of this process, and in
particular of why the gravitational waves emitted by this system are so simple.  Thorne proposed thinking of the zone near each 
rotating black hole as a ``vortex'' in spacetime.  He then proposed that one might be able to understand the gravitational wave
emission by thinking about the interaction of these vorticies with each other.    

Valeria Ferrari spoke on ``Gravitational Waves from Perturbed Stars.''  She noted that she and Chandra had worked on this problem
as an extension of Chandra's earlier work on black hole perturbations.  However, Ferrari pointed out that there are several 
complications that did not arise in the earlier problem because black holes are vacuum solutions, while stars contain matter.  Ferrari
also noted that observations of gravitational waves coming from neutron stars could be used to constrain the neutron star equation
of state.    

At the Banquet there were many stories about Chandra.  My favorite was Eugene Parker's recounting of how his paper on the Solar Wind
came to be accepted for publication in the Astrophysical Journal.  Chandra was editor of Ap. J. and
pointed out to Parker that two referees had recommended against publication.  But Parker replied that the referees had not given any 
good reasons for their opinion, so Chandra decided to publish the paper.  Another excellent story was Virginia Trimble's recounting
of how Chandra replied when asked why he had never been part of any Decadal Survey of Astronomy: with a perfect Yorkshire accent he
quoted a piece of poetry that ended ``Nobody asked me, sir she said.''  

Martin Rees spoke on ``Chandra's Scientific Legacy.''  He pointed out that Chandra had worked on ordinary stars, white dwarfs, 
neutron stars, black holes, and ellipsoids of rotation; all of which are relevant to contemporary astronomy.  Rees then 
concentrated on massive black holes in the centers of galaxies. He noted that observations of the orbits of stars and gas
disks in galactic centers give evidence of the presence of these black holes, as well as measurements of their mass and 
estimates of their spin.  He then mentioned gamma ray bursts, some of which are probably caused by hypernovae that accompany the
formation of a black hole.     

James Stone spoke on ``Magnetohydrodynamics in Astrophysical Contexts.''  He noted that Chandra had done the classic work (and 
written the classic book) on magnetohydrodynamics (MHD).  He then pointed out that MHD is important in many fields of astrophysics, 
and went into some detail on MHD in sunspots and the solar dynamo, accretion disks and their magnetorotational instability, the role
of MHD in jets from black holes, and the role of the magnetic field in the polarization of light scattered from molecular clouds.  

Priyamvada Natarajan spoke on ``The Formation and Growth of Super-Massive Black Holes.''  He pointed out that the standard 
cosmological model of small perturbations and cold dark matter does give rise to regions of high density; but the 
exact process by which this gives rise to supermassive black holes is not completely understood.  It is thought that supermassive black holes start from ``seeds'' (which may be population III stars, or supermassive stars, or the
collapse of a pregalactic disk) and that these seeds form black holes which then grow by a combination of accretion and mergers.   

Ganesan Srinivasan spoke on ``Chandra and the Legacy of Ramanujan.''  He noted that Chandra greatly admired Ramanujan and 
thought of him as a model for great intellectual achievement.  However styles in mathematics differ, and Chandra's style 
was very different from that of Ramanujan.  Thus, Ramanujan was a sort of role model for Chandra, but not one that he tried 
directly to emulate.  

Jeremiah Ostriker spoke on ``Galaxy Structure and Formation.''  He noted that he had tried to emulate Chandra in some ways ({\it e.g.} 
being a workaholic) but not in others ({\it e.g.} elegance of attire).  The subject of his talk was what we learn about galaxies 
from a detailed numerical simulation of their formation and evolution.  The main claim was that large elliptical galaxies are now 
well understood: they accrete other galaxies which makes them so hot that further star formation turns off.  This leads to galaxies 
that are ``red and dead.''

Rashid Sunyaev spoke on ``Scattering of Radiation in the Universe: from the CMB and Last Scattering Surface to Clusters of 
Galaxies and Quasars.''  He noted that Chandra's books were very useful in his scientific research, especially the one on
radiative transfer.  He gave an overview of the history of the universe with emphasis on the cosmic microwave background, and 
the physics of the last scattering surface.  He then talked about the Sunyaev-Zeldovich effect and its use in studying 
galaxy clusters.    

Gordon Garmire spoke on ``The Chandra X-ray Telescope.''  He reviewed the history of the construction and launch of the telescope.  
He then showed some X-ray sources observed by the telescope, including supernova remnants and colliding galaxy clusters.

\end{document}